\documentstyle[emulateapj,psfig]{article}

\begin{document}

\title{The Luminosity--$E_{\rm{p}}$ Relation within Gamma--Ray Bursts and Implications for Fireball Models
}

\author{E. W. Liang$^{1,2,3}$, Z. G. Dai$^{1}$, and X. F. Wu$^{1}$}
\affil{$^1$Astronomy Department, Nanjing University, Nanjing 210093, P. R. China;
Email:ewliang@nju.edu.cn\\ $^2$Physics Department, Guangxi University, Nanning 530004, P. R.
China\\ $^3$National Astronomical Observatories/Yunnan Observatory, Chinese Academy of Sciences,
Kunming 650011, P.R. China}

\begin{abstract}
Using a sample of 2408 time--resolved spectra for 91 BATSE GRBs presented by Preece et al., we show
that the relation between the isotropic--equivalent luminosity ($L_{\rm{iso}}$) and the peak energy
($E^{'}_{\rm{p}}$) of the $\nu F_{\nu}$ spectrum in the cosmological rest frame,
$L_{\rm{iso}}\propto E_{\rm{p}}^{'2}$, holds within these bursts, and also holds among these GRBs,
assuming that the burst rate as a function of redshift is proportional to the star formation rate.
The possible implications of this relation for the fireball models are discussed by defining a
parameter $\omega\equiv (L_{\rm{iso}}/10^{52} {\rm{erg\, s}^{-1}})^{0.5}/(E^{'}_{\rm{p}}/200\,
{\rm{ keV}})$. It is found that $\omega$ is narrowly clustered in $0.1-1$. We constrain some
parameters for both the internal shock and external shock models from the requirement of
$\omega\sim 0.1-1$, assuming that these model parameters are uncorrelated. The distributions of the
parameters suggest that if the prompt gamma--rays are produced from kinetic--energy--dominated
internal shocks, they may be radiated from a region around $R\sim 10^{12}-10^{13}$ cm (or Lorentz
factor $\sim 130-410$) with a combined internal shock parameter $\zeta_{\rm{i}}\sim 0.1-1$ during
the prompt gamma--ray phase, which are consistent with the standard internal shock model; if the
prompt gamma--rays of these GRBs are radiated from magnetic--dissipation--dominated external
shocks, the narrow cluster of $\omega$ requires $\sigma\sim 1-470$, $\Gamma\sim 216-511$,
$E\sim10^{51}-10^{54}$ ergs, $n\sim 0.5-470$ cm$^{-3}$, and $\zeta_{\rm{e}}\sim 0.36-3.6$, where
$\sigma$ is the ratio of the cold-to-hot luminosity components, $\Gamma$ the bulk Lorentz factor of
the fireball, $E$ the total energy release in gamma--ray band, $n$ the medium number density, and
$\zeta_{\rm{e}}$ a combined external shock parameter, which are also in a good agreement with the
fittings to the afterglow data. These results indicate that both the kinetic--energy--dominated
internal shock model and the magnetic--dissipation--dominated external shock model can well
interpret the $L_{\rm{iso}}\propto E_{\rm{p}}^{'2}$ relation and the value of $\omega$.
\end{abstract}

\section{Introduction}               
Gamma--ray bursts (GRBs) are now believed to be produced by jets powered by central engines with a
standard energy reservoir at cosmological distances (see series reviews by Fishman \& Meegan 1995;
Piran 1999; van Paradijs et al. 2000; Cheng \& Lu 2001; M\'{e}sz\'{a}ros 2002; Zhang \&
M\'{e}sz\'{a}ros 2003).

The most impressive features of GRBs are the great diversities of their light curves and spectral
behaviors, and extremely large luminosities. These spectra are well fitted by the Band function
(Band et al. 1993). However, the radiation mechanism at work during the prompt phase remains poorly
understood. Although the spectral behavior and the luminosity are dramatically different from burst
to burst, the isotropic--equivalent luminosity, $L_{\rm{iso}}$, (or isotropic--equivalent energy
radiated by the source, $E_{\rm{iso}}$), and $E^{'}_{\rm{p}}$, the peak energies of $\nu F_{\nu}$
spectrum in the rest frame among GRBs, obey an empirical relation of $L_{\rm{iso}}\propto
E_{\rm{p}}^{'2}$ (Amati et al. 2002; Yonetoku et al. 2003; Sakamoto et al. 2004; Lamb et al. 2003a,
b, c). This relation was revisited in standard synchrotron/inverse--Compton/synchro--Compton models
(Zhang \& M\'{e}sz\'{a}ros 2002). Recently, Sakamoto et al. (2004) and Lamb et al. (2003a, b, c)
pointed out that HETE--2 observations not only confirm this correlation, but also extend it to the
population of X--ray flashes, which are thought to be a low energy extension of typical GRBs (Heise
et al. 2001, Kippen et al. 2003). Based on this relation, Atteia (2003) also constructed a simple
redshift indicator for GRBs.

One may ask: whether or not this relation holds in any segment within a GRB? The answer remains
unknown. If the answer is positive, combining the results mentioned above, one might suggest that
this relation is a universal law during the prompt gamma--ray phase, and presents some constraints
on fireball models. In this Letter, we investigate this issue. Using a sample of 2408
time--resolved spectra for 91 BATSE GRBs presented by Preece et al. (2000), we show that this
relation holds within these bursts, and also holds among these GRBs, assuming that the burst rate
as a function of redshift is proportional to the star formation rate. We suggest that both the
kinetic--energy--dominated internal shock model and the magnetic--dissipation--dominated external
shock model may well interpret this relation.

Throughout this work we adopt $H_0=65$ km s$^{-1}$ Mpc$^{-1}$, $\Omega_{\rm{m}}=0.3$, and
$\Omega_\Lambda=0.7$.

\section{The $L_{\rm{iso}}-E^{'}_{\rm{p}}$ relation within a GRB}
Within a GRB, the relation between $L_{\rm{iso}}$ and $E^{'}_{\rm{p}}$ is equivalent to a relation
between the observed flux ($F$) and peak energy, $E_{\rm{p}}$. Thus, we examine whether or not both
$F$ and $E_{\rm{p}}$ follow a relation of $F\propto E_{\rm{p}}^2$. The time--resolved spectral
catalog presented by Preece et al. (2000) includes 156 long, bright GRBs. Four spectral models were
used in their spectral fittings. Different models might present different fitting results. Among
156 GRBs, 91 GRBs were fitted by the Band function (Band et al. 1993). We only include these GRBs
into our analysis. There are 2408 time--resolved spectra for these GRBs. In our analysis, the
values of $E_{\rm{p}}$ are taken from this catalog. The data used for spectral fittings were
observed by different BATSE detectors. Nominal energy coverage is 25 to 1800 keV, with some small
variations between these detectors. The fluxes presented in the catalog are in an energy band
corresponding to the detectors. Hence, the values of $F$ in our analysis are not the fluxes
presented in the catalog, but are the integrated fluxes in energy band 30--10000 keV (e.g.,
Yonetoku et al. 2003) derived from the model spectral parameters in the catalog.

We evaluate the relation of $F \propto E_{\rm{p}}^2$ within a GRB by the linear correlation
coefficient of the two quantities, log $F$ and log $E_{\rm{p}}^2$. We calculate the linear
correlation coefficients ($r$) and chance probabilities ($p$) for each burst with the Spearman rank
correlation analysis method. The distribution of $r$ is presented in Figure 1. We find that about
75\% GRBs exhibit a strong correlation between the two quantities with $r>0.5$ and $p<0.0001$. We
illustrate 16 cases in Figure 2. These results show that this relation holds within these GRBs.

To examine whether or not this relation holds among these GRBs, we assume that the redshift
distribution for these GRBs is same as that presented by Bloom (2003). Bloom (2003) assumed the
burst rate as a function of redshift is proportional to the star formation rate as a function of
redshift, and presented the observed redshift distribution incorporating with observational biases
(SFR1 model is used in this work, see Porciani \& Madau 2001). We derive a value of redshift for a
given GRB from this distribution by a simple Monte Carlo simulation method. To do so, we first
derive the accumulated probability distribution of the Bloom's redshift distribution, $P(z)$
($0<P(z)\leq 1$), then generate a random number for a given GRB, $m$ ($0<m\leq 1$), and finally
obtain the value of $z$ from $P(z)=m$, i.e., $z=P^{-1}(m)$. Hence, we calculate the values of
$L_{\rm{iso,52}}$ and $E_{\rm{p}}(1+z)$ for these GRBs, where $L_{\rm{iso}, 52}$ is in units of
$10^{52}$ erg s$^{-1}$. The $L_{\rm{iso,52}}$ as a function of $E_{\rm{p}}(1+z)$ is shown in Figure
3. The linear correlation coefficient of the two quantities is 0.63 with a chance probability
$p<0.0001$. A best $\chi^2$ fit with a model of $L_{\rm{iso,52}}\propto [E_{\rm{p}}(1+z)]^2$ is
shown in Figure 3 (solid line). The reduced $\chi^2$ is $0.23$. These results well suggest that the
relation of $L_{\rm{iso,52}}\propto E_{\rm{p}}^{'2}$ holds among these GRBs.

\section{Implications for Fireball Models}
The above results well suggest that the relation of $L_{\rm{iso}}\propto E_{\rm{p}}^{'2}$ remains
within a GRB. This implies that the relation is independent of the temporal evolution of a
fireball. This might provide strong constraints on fireball models. We define a quantity, $\omega$,
to discuss these possible constraints, which is given by

\begin{equation}
\omega=\frac{L_{\rm{iso,52}}^{1/2}}{E^{'}_{\rm{p},200}},
\end{equation}
where $E^{'}_{\rm{p,200}}$ is the peak energy in the rest frame in units of $200$ keV. Since
$L_{\rm{obs}}=\delta ^2L_{\rm{co}}$ and $E_{\rm{obs}}=\delta E_{\rm{co}}$, we obtain
$\omega_{\rm{obs}}=\omega_{\rm{co}}$, where $\delta$ is the Doppler--boosting factor, and the
subscripts ``obs" and ``co" denote the observer frame and the comoving frame, respectively. This
indicates that $\omega$ is not influenced by the Doppler--boosting effect, implying that it seems
to be an intrinsic parameter relevant to the properties of the fireball models.

Two competing fireball models involve the internal shock and external shock models. Zhang \&
M\'{e}sz\'{a}ros (2002) analyzed correlations between $E_{\rm{p}}$ and other parameters in several
cases. From their results, we find that the kinetic--energy--dominated internal shock model (low
$\sigma$, where $\sigma$ is the ratio of the cold-to-hot luminosity components) and
magnetic--dissipation--dominated external shock model (high $\sigma$) have a potential to interpret
the $L_{\rm{iso}}\propto E^{'2}$ relation. From their Eqs. (17) and (18), we derive
\begin{equation}
\omega\simeq \zeta_{\rm{i}} R_{13}
\end{equation}
for the internal shock model with low $\sigma$, where $\zeta_{\rm{i}}$ is a combined internal shock
parameter\footnote{$\zeta_{\rm{i}}=\epsilon^{-1}_{\rm{x3}}$, see Zhang \& M\'{e}sz\'{a}ros
(2002).}, and $R_{13}$ the radius of the fireball in units of $10^{13}$ cm. From their Eq
(23)\footnote{We check Eq. (23) in Zhang \& M\'{e}sz\'{a}ros (2002) and find that the coefficient
of this equation is 12 keV, rather than 880 keV. In order to discuss conveniently the distribution
of $\sigma$, we also assume $\sigma^{1/2}/(1+\sigma)^{1/6}\simeq \sigma^{1/3}$. Thus, the equation
is re--scaled as $E_{\rm{p}}^e(high \ \sigma)\simeq 260 {\rm{keV}} \zeta_{\rm{e}} \sigma_{1}^{1/3}
\Gamma_{2.5}^{8/3} L_{52}^{1/2} E_{53}^{-1/3} n_{1}^{1/3}(1+z)^{-1}$,  where $\zeta_{\rm{e}}$ is a
combined external shock parameter
($\zeta_{\rm{e}}=\epsilon^{-1}_{\rm{w,-1}}\epsilon_{\gamma}^{-2}\sin^{-1}\Psi$, see Zhang \&
M\'{e}sz\'{a}ros 2002), $\sigma_{1}=\sigma/10$, $\Gamma_{2.5}=\Gamma/10^{2.5}$, $E_{53}$ the total
energy in units of $10^{53}$ ergs, and $n_{1}$ the medium number density in units of $10$
cm$^{-3}$.}, we obtain

\begin{equation}
\omega\simeq 0.36 \zeta_{\rm{e}} \sigma_{1}^{-1/3} \Gamma_{2.5}^{-8/3}  E_{53}^{1/3} n_{1}^{-1/3}
\end{equation}
for the external shock model with high $\sigma$.

From Eq. (2) one can see that, for the internal shock model, $\omega$ is determined by the radius
of the  gamma--ray emission region (hence the bulk Lorentz factor since $R\propto \Gamma^2$) and
$\zeta_{\rm{i}}$. Please note that $\zeta_{\rm{i}}$ is related to the shock parameters, such as the
index of the electron distribution, the electron equipartition factor, the magnetic equipartition
factor, the pitch angle of electrons, the relative Lorentz factor between the shells, etc. These
parameters are only related to the physics of colliding shells. However, for the external shock
model, the case becomes more complicated. Eq. (3) shows that $\omega$ is determined by the
parameters of both the shocks and surroundings.

We calculate $\omega$ for each temporal segment within a GRB and investigate the evolution of
$\omega$ by a linear model of $\omega \propto k t$. The value of $k$ evaluates the general trend of
the temporal evolution feature of $\omega$: the larger the absolute value of $k$ is, the more
significantly $\omega$ evolves. The distributions of  $\omega$ and $k$ are shown in Figures 4 and
5, respectively. We find that $\omega$ mainly distributes in $0.1-1$, and $k$ is narrowly clustered
in -$0.03-0.03$. These results show that $\omega$ seems to be an invariant without temporal
evolution for different GRBs and even for different temporal segments within a GRB .

The parameters in Eqs. (2) and (3) seem to be uncorrelated, although we do not know if this is
really the case. We simply assume that they are uncorrelated, and constrain their distributions
from the requirement of $\omega\sim 0.1-1$. We suggest that these parameters should be clustered in
the same range as that of $\omega$. Thus, we derive $R\sim10^{12}-10^{13}$ cm and
$\zeta_{\rm{i}}\sim 0.1-1$ for the internal shock model, implying that most of the gamma--rays are
radiated from a region around $R\sim 10^{12}-10^{13}$ cm with similar shock--acceleration and
radiation mechanisms during the prompt gamma--ray phase. Since $R\simeq 2\Gamma ^2 c \delta
t_{\rm{v}}\sim 0.6\times 10^{13} \Gamma_{2.5}^2 \delta t_{\rm{v, -3}} \rm{cm}$ , where c is the
speed of light, and $t_{\rm{v,-3}}$ the variability timescale in units of $10^{-3}$ second, we
obtain $\Gamma\sim 130-410$. These parameters are consistent with the standard internal shock
model. For the external shock model, we derive $\sigma \sim 1-470$, $\Gamma\sim 216-511$,
$\zeta_{\rm{e}}\sim 0.36- 3.6$, $E\sim 10^{51}-10^{54}$ ergs, and $n\sim 0.5-470$ cm$^{-3}$. The
distributions of these parameters are in a good agreement with the fittings to the afterglows
(Panaitescu \& Kumar 2001).

The above results indicate that both the low $\sigma$ internal shock model and the high $\sigma$
external shock model can well interpret the $L\propto E_{\rm{p}}^{'2}$ relation and the value of
$\omega$.

\section{Conclusions and Discussion}
Using a sample of 2408 time--resolved spectra for 91 long, bright GRBs presented by Preece et al.
(2000), we show that the $L_{\rm{iso}}\sim E_{\rm{p}}^{'2}$ relation holds within these BATSE
bursts, and this relation also holds among these GRBs by assuming that the burst rate as a function
of redshift is proportional to the star formation rate.

We discuss possible implications of this relationship for the fireball models by defining a
parameter $\omega\equiv (L_{\rm{iso}}/10^{52} {\rm{erg\ s}^{-1}})^{0.5}/(E^{'}_{\rm{p}}/200\,
{\rm{keV}})$. It is found that $\omega$ is not influenced by the Doppler--boosting effect, and it
is determined by the gamma--ray emission region and shock parameters in the
kinetic--energy--dominated internal shock model or determined by the parameters of both the shock
and the environment in the magnetic--dissipation--dominated external shock model. We derive the
distributions of some parameters for both the internal shock model and the external shock model
from the requirement of $\omega\sim 0.1-1$. We suggest that if the prompt gamma--rays are produced
from a kinetic--energy--dominated internal shock, they may be radiated from a region around $R\sim
10^{12}-10^{13}$ cm (or Lorentz factor $\sim 130-410$) with an internal shock parameter
$\zeta_{\rm{i}}\sim 0.1-1$, which is consistent with the standard internal shock model; if the
prompt gamma--rays of these GRBs are radiated from magnetic--dissipation--dominated external
shocks, the $\omega\sim 0.1-1$ requires $\sigma \sim 1-470$, $\Gamma \sim 216-511$,
$\zeta_{\rm{e}}\sim 0.36-3.6$, $E\sim 10^{51}-10^{54}$ ergs, and $n\sim 0.5-470$ cm$^{-3}$. Please
note that the distributions for these model parameters for both the internal and external shock
models are based on the assumption that they are uncorrelated. Although these parameters seem to be
uncorrelated, we do not know if it is really the case. If these parameters are correlated during
prompt gamma--ray phase, these distributions are not valid.

We would like to thank the referee, Dr. Don Lamb, for his valuable comments, which have enabled us
to improve greatly the manuscript. This work is supported by the National Natural Science
Foundation of China (grants 10233010 and 10221001), the National 973 Project (NKBRSF G19990754),
the Natural Science Foundation of Yunnan (2001A0025Q), and the Research Foundation of Guangxi
University.

\begin{figure}
\plotone{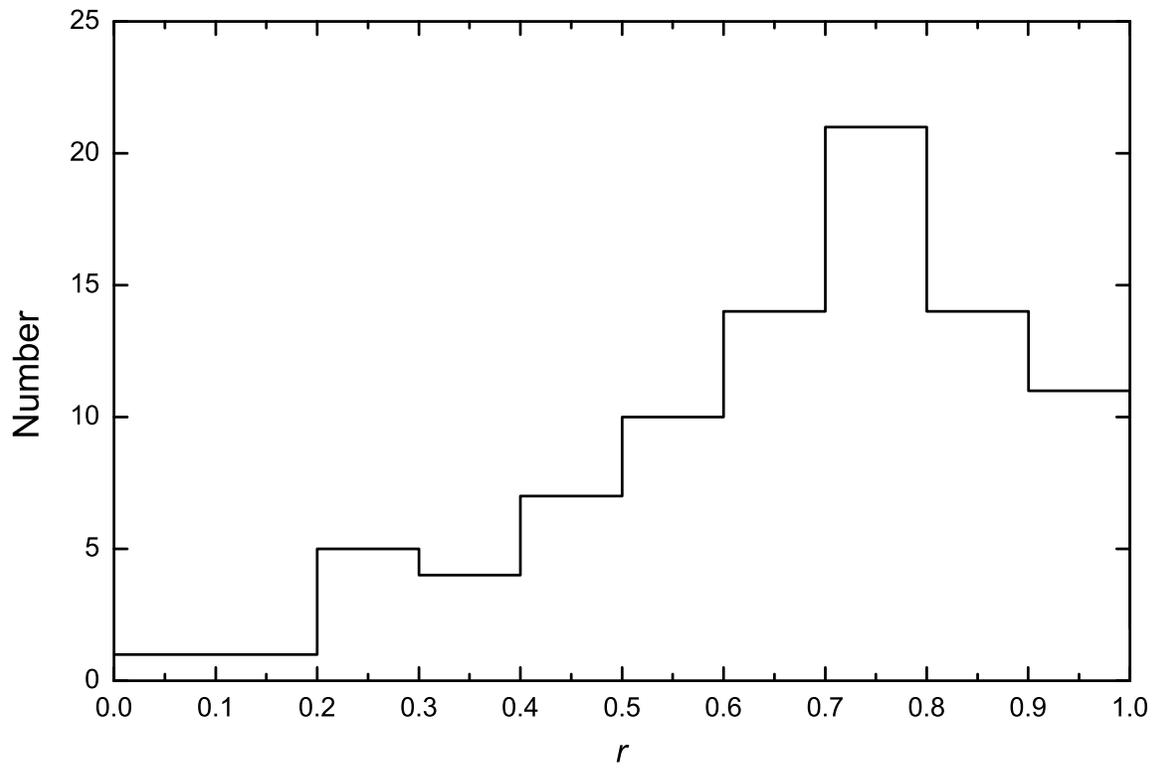} \caption{Distribution of the linear coefficients for log$F$--log$E_{\rm{p}}^2$.
\label{fig1}}
\end{figure}

\begin{figure}
\plotone{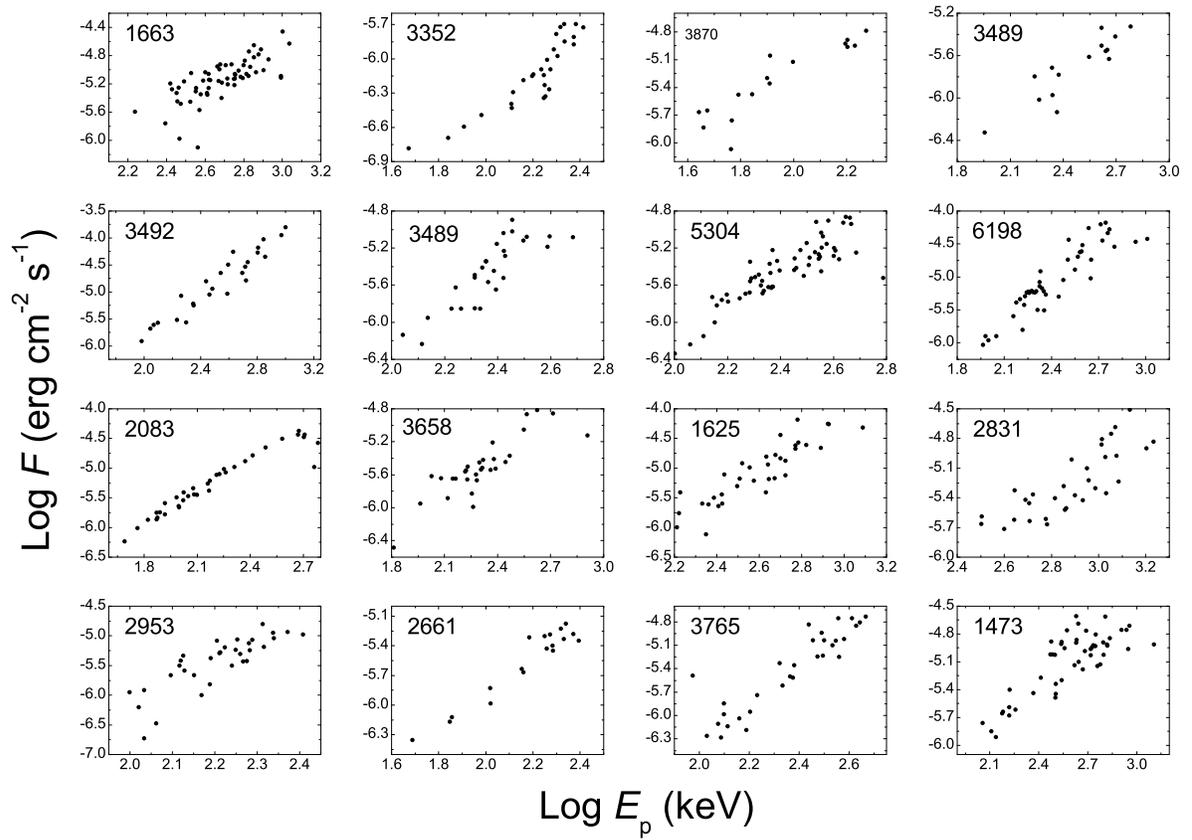}\caption{The observed flux as a function of $E_{\rm{p}}$ for 16 GRBs.
\label{fig2}}
\end{figure}

\begin{figure}
\plotone{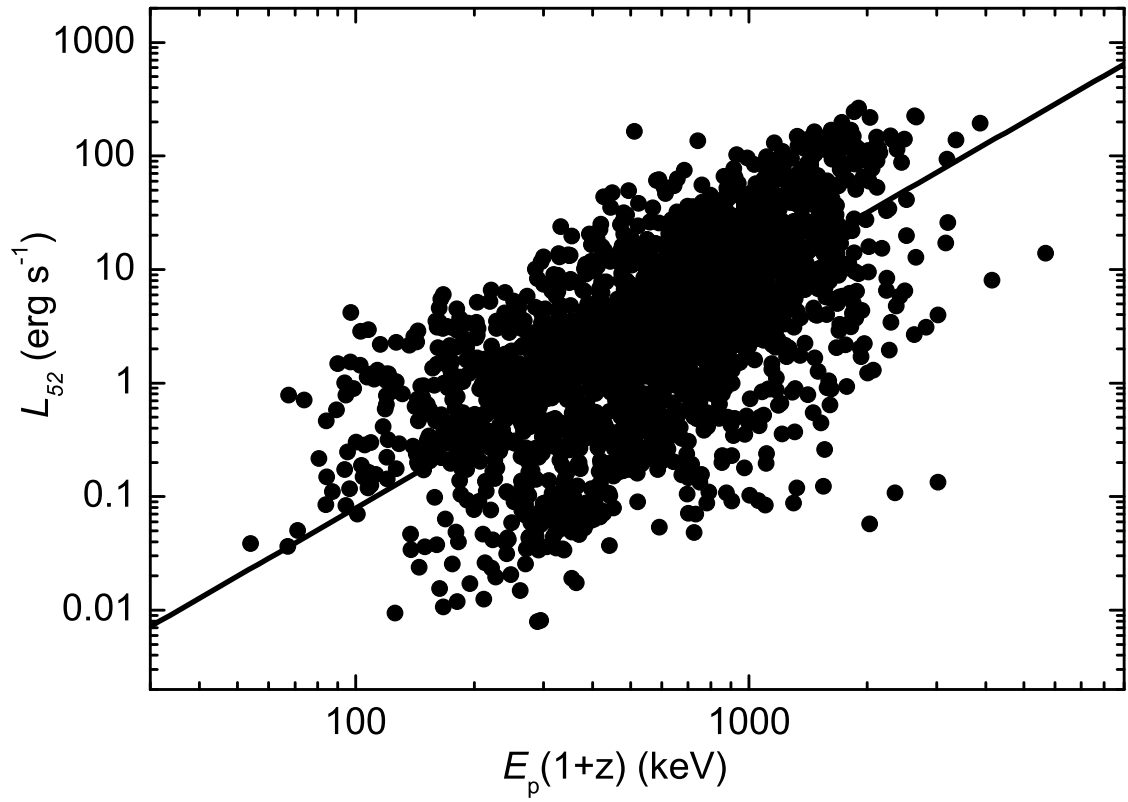} \caption{The $L_{\rm{iso,52}}$ as a function of $E_{\rm{p}(1+z)}$ for 2408 GRB
spectra. The solid line is $L_{\rm{iso, 52}}=10^{-5.1}\times [E_{\rm{p}}(1+z)]^2$. \label{fig3}}
\end{figure}

\begin{figure}
\plotone{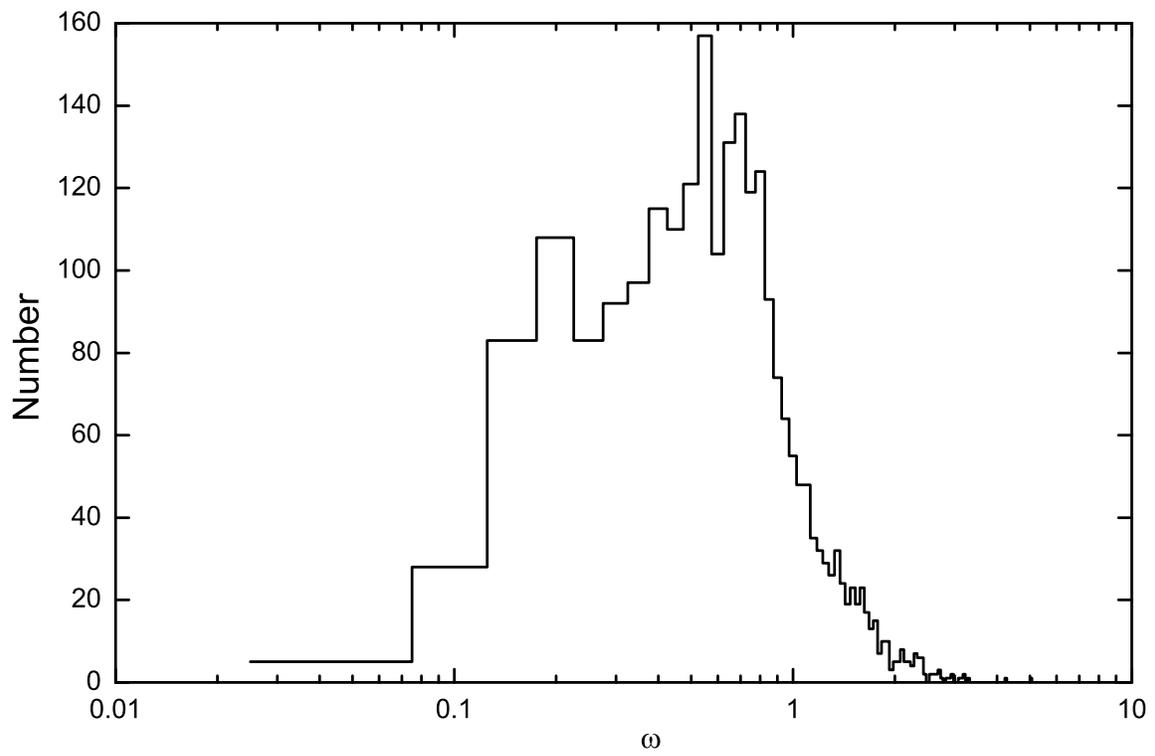} \caption{Distribution of $\omega$ for 2408 spectra. \label{fig4}}
\end{figure}

\begin{figure}
\plotone{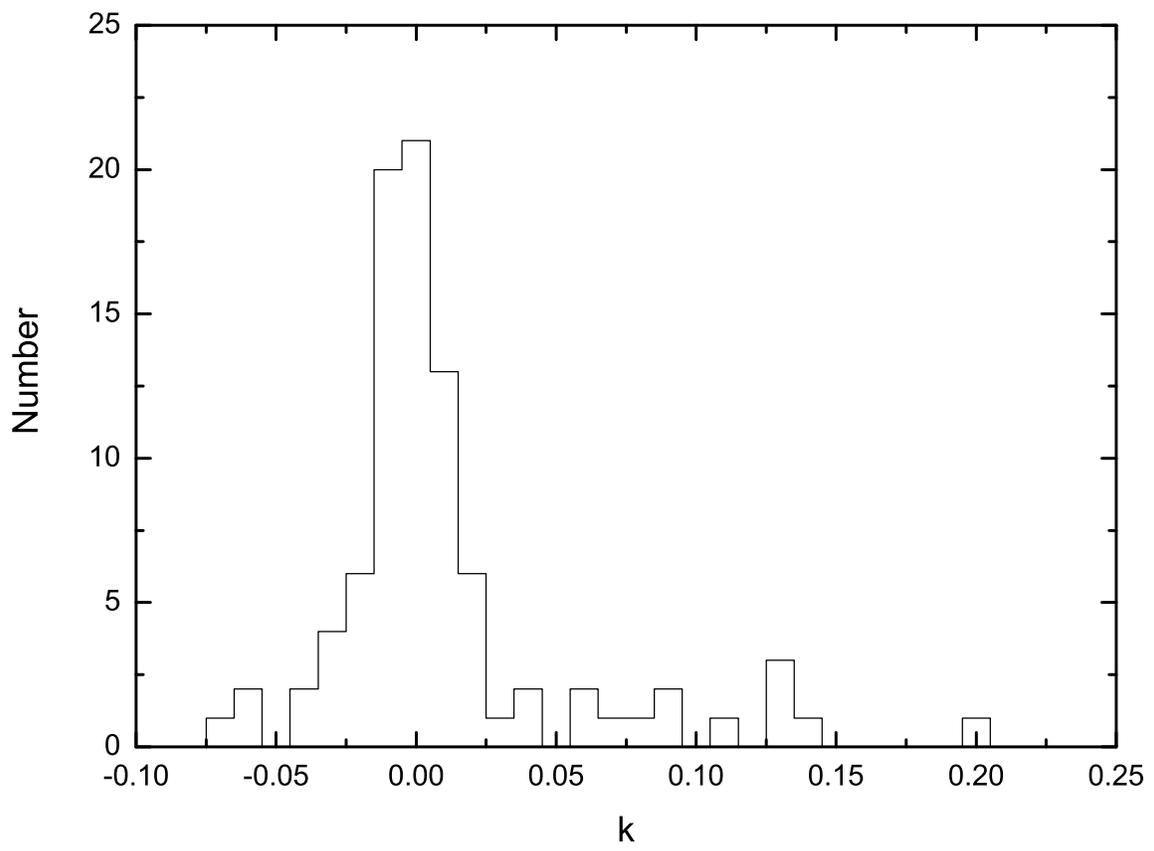} \caption{Distribution of $k$ for 91 GRBs. \label{fig5}}
\end{figure}

%
%
%
%
%
%
%

\end{document}